# Surface Passivation for Halide Optoelectronics: Comparing Optimization and Reactivity of Amino-Silanes with Formamidinium

*Zixu Huang, [1,†] Farhad Akrami,[1,†] Junxiang Zhang,[2] Stephen Barlow,[2] Seth R. Marder,[2,3] David S. Ginger[1]\**

[1]Department of Chemistry, University of Washington, Seattle, WA 98195, USA

[2]Renewable and Sustainable Energy Institute (RASEI), University of Colorado Boulder, Boulder, Colorado 80309, USA

[3]Departments of Chemical and Biochemical Engineering and of Chemistry, University of Colorado Boulder, Boulder, Colorado 80309, USA

†These authors contributed equally to this work.

*Corresponding author: dginger@uw.edu

Amino-silane-based surface passivation schemes are gaining attention in halide perovskite optoelectronics, with varying levels of success. We compare surface treatments using (3-aminopropyl)trimethoxysilane (APTMS) and [3-(2-aminoethylamino)propyl]trimethoxysilane (AEAPTMS), applied via room-temperature vacuum deposition, to the perovskite $FA_{0.78}Cs_{0.22}Pb(I_{0.85}Br_{0.15})_3$ (FA = formamidinium). Both molecules improve thin-film photoluminescence properties and photovoltaic device performance, although their effectiveness depends strongly on deposition time. We show AEAPTMS has a wider, more robust processing window and yields higher performance under optimized conditions. In contrast, over-exposure, particularly with APTMS, reduces performance, with notable reductions in photoluminescence lifetime and absorbance. To probe the underlying chemistry, we employ nuclear magnetic resonance (NMR) spectroscopy and depth-resolved time-of-flight secondary ion mass spectrometry (ToF-SIMS), demonstrating that both amino-silanes react with formamidinium ($FA^+$) cations in solution and in the solid state. This work underscores the importance of optimizing deposition conditions to balance effective passivation with potential performance loss and elucidates previously unrecognized reactive chemistry between amino-silane passivating agents and halide perovskites.

**TOC Graphic**

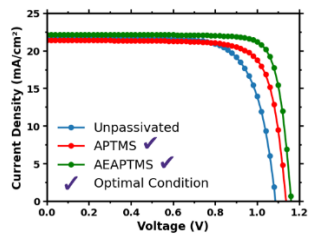

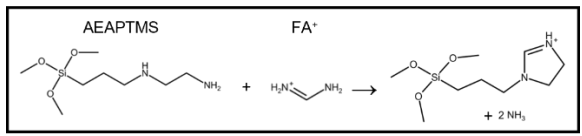

Metal halide perovskites are semiconductors with the general formula of $ABX_3$ where A is a monovalent cation (typically methylammonium ($MA^+$), formamidinium ($FA^+$), or $Cs^+$), B is a divalent cation (typically $Pb^{2+}$ or $Sn^{2+}$), and X is a halide anion ($I^-$, $Br^-$, or $Cl^-$).[1,2] These materials have attracted much attention for their versatility and performance across a wide range of applications, including photovoltaics, light-emitting diodes, photodetectors, and even single photon sources.[3–11]

While halide perovskites are widely described as defect tolerant, like all semiconductors they exhibit a variety of bulk and surface defects.[12–16] In perovskites these defects contribute to detrimental processes ranging from non-radiative recombination[15,17] to ion migration, phase-separation, and instability.[3,18–23] Mitigating these defects, particularly unpassivated surface states, is essential for enhancing the performance of perovskite semiconductors in applications from tandem solar cells to light-emitting applications.[13,21,24–27] Furthermore removing such defects can significantly enhance both material and device stability.

Numerous studies have demonstrated effective surface passivation strategies that reduce nonradiative recombination in perovskite films and enhance solar cell performance.[15,17,28–32] Notably, studies report that amino-silanes are effective as scalable surface passivators in both films and complete devices, resulting in enhanced photoluminescence properties[15,33] and corresponding improvements in open-circuit voltage,[28,33] suppression of voltage-induced ion migration,[20] decreased photoinduced halide segregation,[19] and even improved device stability under full-spectrum sunlight at 85 °C and open-circuit conditions in ambient air.[33]

While early studies often focused on (3-aminopropyl)trimethoxysilane (APTMS), Lin et al. recently compared a series of amine-functionalized silanes, reporting that [3-(2-aminoethylamino)propyl]trimethoxysilane (AEAPTMS) yielded superior device performance and stability. While APTMS contains a single primary amine, AEAPTMS contains neighboring primary and secondary amines. Based on density functional theory (DFT) calculations, Lin et al. proposed that AEAPTMS passivates more favorably by binding to undercoordinated $Pb^{2+}$ cations in a cooperative fashion. Interestingly, they also reported that APTMS passivation decreased cell performance, in apparent contradiction to a number of previous reports.[15,19,20,28,34]

In this work, we study the effect of time-optimized, vapor-deposited surface treatments with APTMS and AEAPTMS on a $FA_{0.78}Cs_{0.22}Pb(I_{0.85}Br_{0.15})_3$ perovskite with a bandgap of 1.66 eV.

We used this perovskite formulation due to its reported stability and its suitability as a top absorber layer in perovskite–silicon tandem solar cells.[3,35–38] Using time-resolved photoluminescence (TRPL), we show that both APTMS and AEAPTMS treatments increase carrier lifetimes, as well as open-circuit voltage and power-conversion efficiency of solar cells. We show that the overall effectiveness of the passivation varies with deposition time (silane coating thickness) and that the optimal conditions differ not only between APTMS and AEPTMS but also depending on if we optimize for maximum PL lifetime, or maximum device performance. Finally, we show that both APTMS and AEAPTMS react with $FA^+$ cations, challenging some current models for surface binding.[33]

We synthesized $FA_{0.78}Cs_{0.22}Pb(I_{0.85}Br_{0.15})_3$ perovskites by adapting previously reported procedures,[35] (see Supporting Information (SI) for details). We measure a bandgap of ~1.66 eV via UV-Vis spectroscopy (Figure S1). For all surface treatments, we deposit both APTMS and AEAPTMS using low pressure vacuum deposition (see SI). Vapor deposition at reduced pressure is scalable, favors deposition of the silane monomer, and offers improved reproducibility.[39] We also find it reduces pooling of liquid silane derivative on the perovskite surface, thus reducing the prospects for etching. Figure 1a shows the molecular structures of APTMS and AEAPTMS.

First, we investigate the effects of time-dependent APTMS and AEAPTMS treatments on TRPL and UV-Vis absorbance (Figure 1b-e). Figures 1b and 1c show that for deposition times of 30 s, 90 s, and 360 s, at room temperature with gauge pressure of about -25 in. of Hg relative to atmospheric pressure, both APTMS and AEAPTMS treatments extend TRPL lifetimes compared to the unpassivated sample. Table S1 summarizes fitting parameters for the stretched exponential decay of TRPL data.

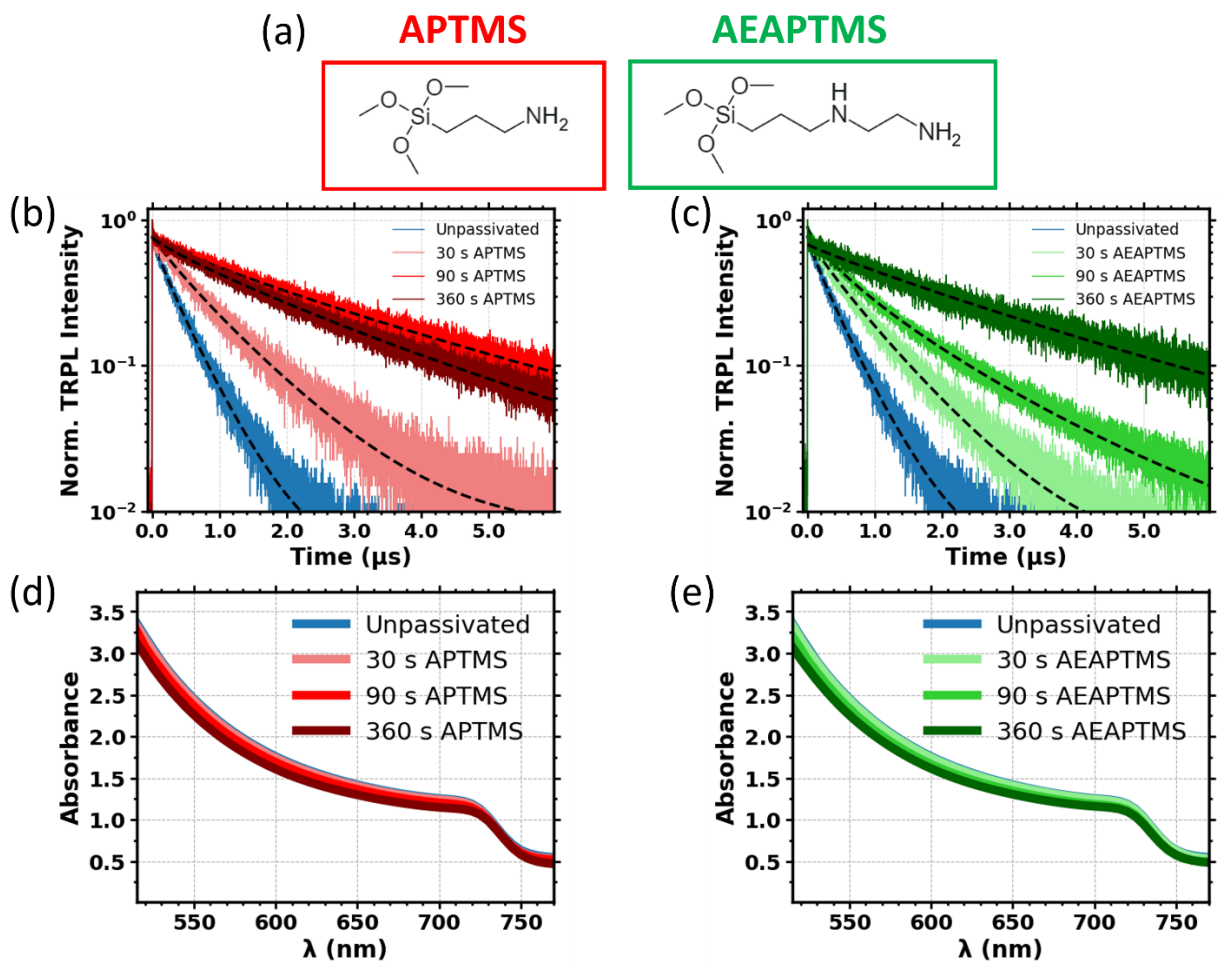

Figure 1. Deposition time-dependent surface treatment of $FA_{0.78}Cs_{0.22}Pb(I_{0.85}Br_{0.15})_3$ perovskites with APTMS and AEAPTMS. (a) Molecular structures of APTMS (red) and AEAPTMS (green). (b) and (c) TRPL decay curves of unpassivated (blue), APTMS-passivated (red), and AEAPTMS-passivated (green) samples with varying deposition times. Black dashed lines are stretched exponential fits to the data. (d) and (e) UV-Vis absorbance spectra of unpassivated (blue), APTMS-passivated (red), and AEAPTMS-passivated (green) samples with varying deposition times.

We observe that longer amino-silane deposition times generally result in successively longer TRPL lifetimes, except for the longest (360 s) APTMS treatment, which does not show an increased lifetime relative to the 90 s APTMS treatment. Notably, we achieve similar TRPL enhancements with both APTMS and AEAPTMS, though at different optimum deposition times (see Table S1). Thus UV-Vis spectra in Figure 1d and 1e show a small but measurable decline in extinction with increasing deposition time for both studied silanes. The X-ray diffraction (XRD) patterns (Figure S2) of the untreated and treated samples across the same deposition time range show a subtle reduction in peak intensities at higher treatment times. Collectively, these results indicate that both APTMS and AEAPTMS can effectively passivate perovskite surface defects, though a trade-off exists between passivation and possible disruption of the perovskite structure that is sensitive to silane-dependent deposition conditions.

Next, we examine the incorporation of APTMS and AEAPTMS surface treatments into archetypal p–i–n structured perovskite solar cells. Figure 2a shows the detailed schematic of the device architecture. This commonly used architecture comprises an ITO electrode treated with [2-(3,6-dimethoxy-9H-carbazol-9-yl)ethyl]phosphonic acid (MeO-2PACz) as the hole-transport layer, evaporated $C_{60}$ as the electron-transport layer, and Ag as the back electrode. Each of these layers is currently widely used and known to yield well-performing devices.[37,40–43]

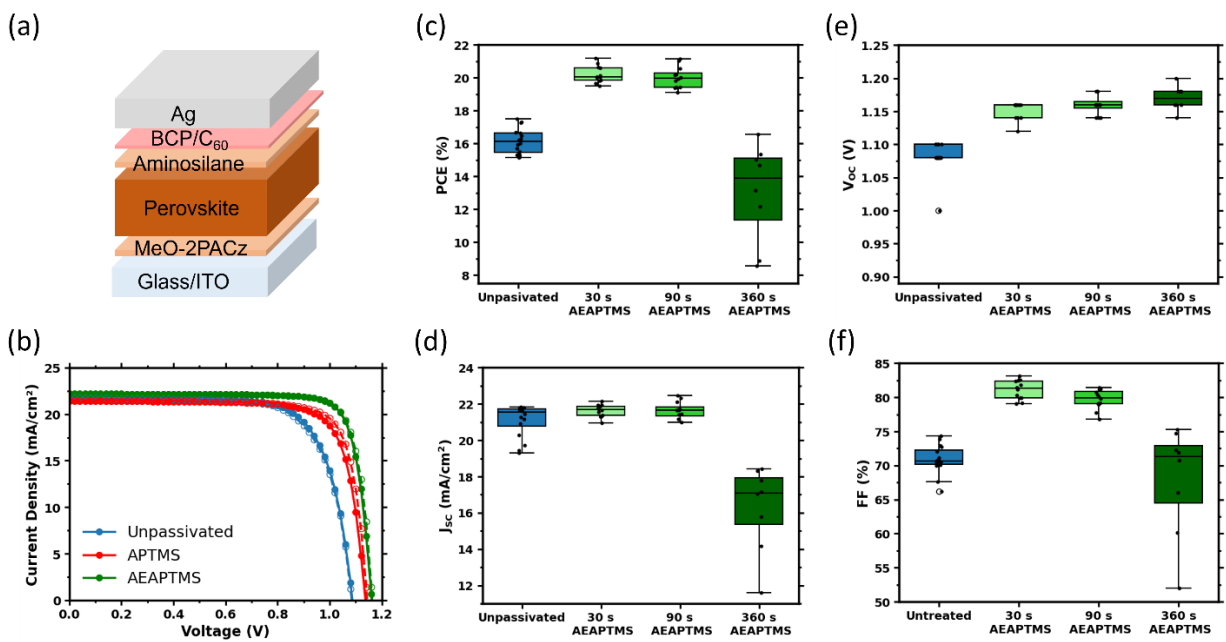

Figure 2. Perovskite solar-cell performance with deposition time-dependent amino-silane surface treatments. (a) A schematic illustration of the p–i–n structured perovskite solar cell. (b) J-V curves of the champion unpassivated (blue), APTMS-treated (red), and AEAPTMS-treated (green) devices with 30 s treatment duration. Solid lines represent forward scans and dotted lines represent reverse scans. (c) Power conversion efficiency (PCE), (d) short-circuit current density ($J_{SC}$), (e)

open-circuit voltage ($V_{OC}$), and (f) fill factor (FF) of unpassivated (blue) and AEAPTMS-treated (green) devices with varying deposition times (based on forward scans).

We observe that longer amino-silane deposition times eventually lead to a decline in solar cell performance (Figures 2c–f and S3), and we find the optimal performance (Figure 2b) occurs with shorter treatment durations (~30 s). Under optimized treatment conditions, both APTMS- and AEAPTMS-treated devices show improved power conversion efficiencies, most notably through enhanced open-circuit voltage ($V_{OC}$) and fill factor (FF). Among the two treatments, AEAPTMS-treated solar cells demonstrate the highest overall efficiency. Table 2 summarizes the mean values and standard deviations of the device data. It is noteworthy that, consistent with the trend of increased TRPL lifetimes at longer amino-silane deposition times (Figure 1b-c), the $V_{OC}$ also tends to improve with extended treatment durations. This result is widely consistent with past data on amino-silane treatment of perovskites.[28,33]

However, despite the $V_{OC}$ improvements, we observe a decrease in $J_{SC}$, FF, and PCE at longer deposition times. We attribute this trade-off to a combination of perovskite decomposition and, more significantly, the insulating nature of the amino-silanes,[28] which becomes more detrimental as thicker layers provide a greater barrier to charge extraction after prolonged deposition. Furthermore, at even longer deposition times (Figure S4), or under more aggressive conditions,[33] some reduction in the 3D perovskite phase occurs as evidenced by the concomitant decrease in UV-Vis absorbance and XRD peak intensities. In our system, we observe a similar decrease in the XRD peak intensities at 900 s (Figure S4), supporting this loss trend. Overall, these findings highlight a tunable balance between beneficial surface modification and adverse effects, namely electrical insulation and decomposition, which is sensitive to treatment duration and conditions.

On one hand, these observations are consistent with previous work assigning the beneficial effects of amino-silane treatment to the passivation of surface defects, through processes such as coordination of the Lewis basic amine groups to $Pb^{2+}$ sites with surface halide vacancies.[15,28,33] On the other hand, our group and others have reported enhanced performance in amine-based passivation strategies where we have conclusively demonstrated that the amine groups react with the $FA^+$ cations.[44–46] Therefore, to better understand how these two amino-silanes interact differently with the perovskite, we monitored the interaction between AEAPTMS and FAI in DMSO-$d_6$ (0.04 M) by solution $^1H$ NMR spectroscopy.

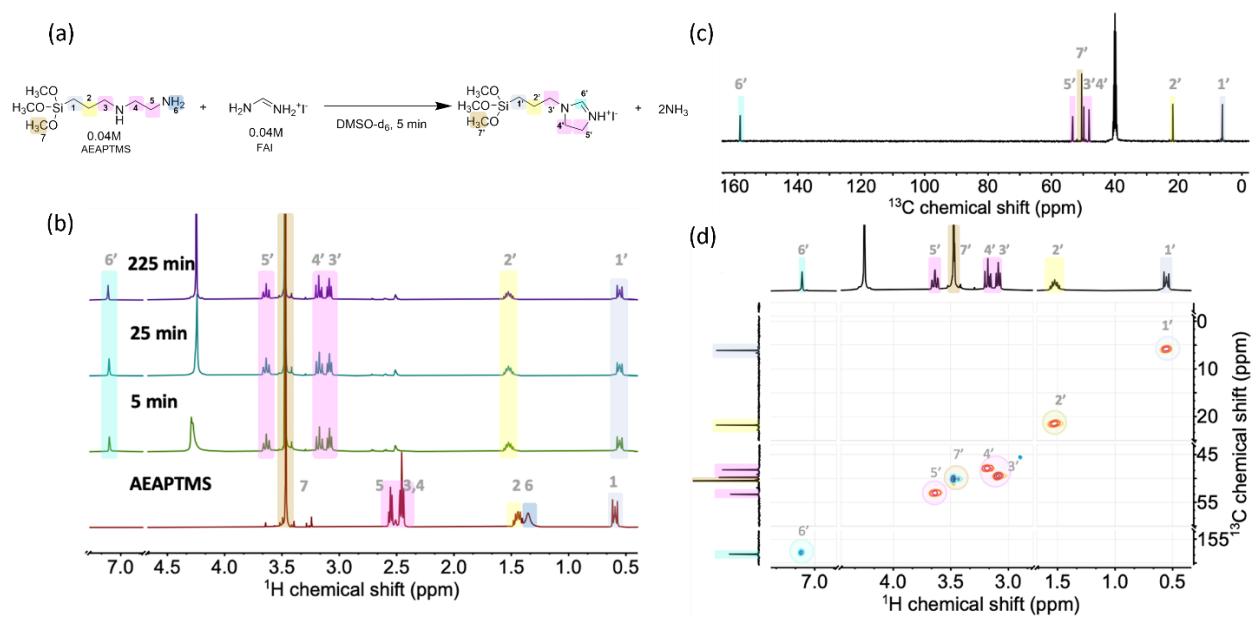

Figure 3. (a) Overall reaction between AEAPTMS and FAI, (b) $^1$H NMR spectra of AEAPTMS in DMSO-$d_6$ (0.08 M), and the spectrum of AEAPTMS and FAI mixture after mixing (0.04 M, 1:1 ratio) in DMSO-$d_6$ for 5, 15, and 225 min, indicating that the cyclization reaction is essentially complete in ca. 5 min. (c) $^{13}$C{$^1$H} NMR spectrum and (d) $^1$H-$^{13}$C Heteronuclear Single Quantum Coherence (HSQC) NMR spectrum of the AEAPTMS and FAI mixture after mixing (0.04 M, 1/1 ratio) in DMSO-$d_6$ for 10 h. The relatively clean $^{13}$C NMR spectrum suggests the cyclization reaction is complete and the product is stable in DMSO-$d_6$.

Figure 3b shows $^1$H NMR spectra of AEAPTMS and the mixed solutions of AEAPTMS and FAI in DMSO-$d_6$ (0.04 M) with varying mixing times. Within 5 minutes, the mixed solution shows a new peak in the $^1$H NMR spectrum at a chemical shift of 7.10 ppm, while the characteristic methylene resonances of AEAPTMS, especially those assigned to CH$_2$N groups, shift and the signal corresponding to the methine resonance of FA$^+$ (7.86 ppm)[46] is absent. As shown in Figure 3c, the $^{13}$C{$^1$H} NMR spectrum of the same solution after mixing for 10 hours shows a new signal at 158.1 ppm, indicating the formation of a new sp$^2$ carbon environment not attributable to FA$^+$. Heteronuclear single quantum correlation (HSQC) spectroscopy (Figure 3d) indicates that the proton and carbon associated with these two new resonances are bonded to one another. This data, along with Figures S7–S8, suggest the nucleophilic attack of AEAPTMS on FA$^+$, leading to ring formation with concomitant elimination of ammonia. Figure S9 shows the mass spectrum of AEAPTMS and mixed solution of AEAPTMS and FAI. The signal at $m/z$ = 233.3 is also consistent with the formation of 1-(3-(trimethoxysilyl)propyl)-4,5-dihydro-1*H*-imidazol-3-ium cation after mixing AEAPTMS and FAI, thus strongly supporting the proposed reaction.

In contrast, the reaction between APTMS and FAI in DMSO-$d_6$ (0.04 M) appears to be ca. 50% complete after 5 minutes and does not proceed further, even over many hours (Figure S10a). The expected initial products of APTMS and FA$^+$ are the *N*-(3-(trimethoxysilyl)propyl)formamidinium ion and ammonia. Subsequent proton transfer from *N*-(3-(trimethoxysilyl)propyl)formamidinium to a second molecule of APTMS would result in the formation of a *N*-(3-(trimethoxysilyl)propyl)formamidine and the non-nucleophilic 3-(trimethoxysilyl)propylammonium ion. Although typical formamidinium derivatives are insufficiently acidic to protonate typical primary amines, in the present case we hypothesize that the formamidine product of this proton-transfer equilibrium is stabilized by coordination of the deprotonated nitrogen to the Lewis acidic silicon center, forming a five-membered ring (Figure S10b, S11). The reaction *does* proceed to completion if conducted in methanol-$d_4$ (Figure S12), which can compete for binding with the silicon center (as evidenced by scrambling of $CD_3$ groups into the $Si(OMe)_3$ group, SI Figure S13), or in the presence of the strong non-nucleophilic base DBU, which accepts the proton generated on forming the internal formamidine-Si adduct (Figure S11, S14). Similar interactions are possible in the case of the AEAPTMS / FA reaction; however, a five-membered cyclic internal formamidine adduct analogous to that proposed above for the APTMS cannot be formed since there is no appropriate NH proton to be lost, a possible eight-membered adduct is likely less stable, and, critically, the formation of the cyclic 1-(3-(trimethoxysilyl)propyl)-4,5-dihydro-1*H*-imidazol-3-ium cation is thermodynamically favorable. We also use $^1$H, $^{13}$C{$^1$H}, and 2D NMR methods, as well as mass spectrometry (MS), to confirm the structures of reaction products (Figure S18). Mass spectroscopy suggests reaction of AEAPTMS and FAI exclusively forms of the above-mentioned imidazolium derivative, with no detectable uncyclized products (Figure S9). In contrast, mass spectra for the reaction of APTMS and FAI (1:1) in methanol show peaks assignable to both *N*-(3-(trimethoxysilyl)propyl)formamidinium (*m/z* = 207.2) and *N,N*′-bis(3-(trimethoxysilyl)propyl)formamidinium (*m/z* = 369.2) (Figure S18).

Given the difference in reactivity of formamidinium with amino-silanes in solution, we employed time-of-flight secondary ion mass spectroscopy (ToF-SIMS) to explore the interaction of the amino-silanes and FA$^+$ on solid-state perovskite interfaces.

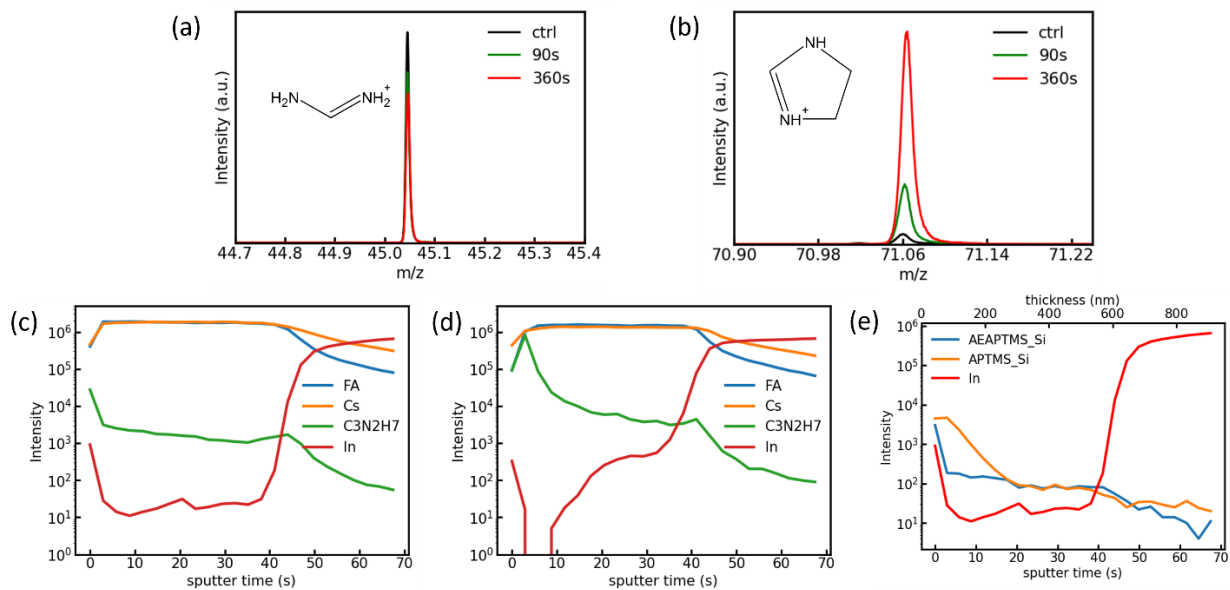

Figure 4. ToF-SIMS mass spectra at (a) m/z = 45.05 and (b) m/z = 71.06, corresponding to FA$^+$ and 4,5-dihydroimidazolium respectively, of perovskite films passivated with AEAPTMS using different deposition times. ToF-SIMS depth profile of (c) unpassivated and (d) AEAPTMS-passivated film (360 s) on ITO substrate. (e) Si depth profile of perovskite films on ITO substrate with different passivation at 90 s.

Figure 4a-b shows the signal of FA$^+$ (m/z = 45.05) and 4,5-dihydroimidazolium ($C_3H_2N_7^+$, m/z = 71.06, a fragment of 1-(3-(trimethoxysilyl)propyl)-4,5-dihydro-1$H$-imidazol-3-ium) respectively, from ToF-SIMS mass spectra of AEAPTMS-passivated perovskite films with varying deposition times. Figure 4a shows the signal of FA$^+$ decreases continuously with increasing duration of AEAPTMS-treatment, while Figure 4b shows a concomitant increase in the signal of $C_3H_2N_7^+$ (4,5-dihydroimidazolium), as expected if the AEAPTMS reacts with FA$^+$ to form 1-(3-(trimethoxysilyl)propyl)-4,5-dihydro-1$H$-imidazol-3-ium as suggested by the solution NMR (Fig 3a). The full mass spectrum also shows a weak signal (likely due to fragmentation) at $m/z$ = 233.3 corresponding to the product of AEAPTMS-FA$^+$ reaction (1-(3-(trimethoxysilyl)propyl)-4,5-dihydro-1$H$-imidazol-3-ium), again in good agreement with the solution reaction scheme. We do detect a small signal at an $m/z$ that is indicative of $C_3H_2N_7^+$ for an unpassivated film, which we speculate could arise from ion bombardment-induced fragmentation and subsequent recombination of $FA^+$ (Fig. 4c). Nevertheless, the signal associated with $C_3H_2N_7^+$ ring increases by ~20 times after exposure to AEAPTMS (Fig. 4d). Together, these results confirm that the reaction of AEAPTMS and $FA^+$ takes place rapidly on the perovskite surface, with increasing product formed at increasing deposition times.

Next, we examine the ToF-SIMs results on APTMS-treated perovskite samples, which show no evidence for *N*-(3-(trimethoxysilyl)propyl)formamidium (*m/z* = 207.2) or *N,N'*-bis(3-(trimethoxysilyl)propyl)formamidinium (*m/z* = 369.2) mass fragments in films. In addition, for APTMS, we observe no significant changes in the signal of $FA^+$ with deposition time (Figure S19), which is in contrast with the decrease in $FA^+$ signal with increasing deposition time for AEAPTMS (Fig. 4a). It is possible that the high fragmentation probabilities of the C-N bonds would lead to dissociation of many of these compounds into APTMS and $FA^+$ like fragments. However given that we see (weak) signals for the parent fragment of AEAPTMS reacting with $FA^+$, we conclude that while it is possible that 1-(3-(trimethoxysilyl)propyl)formamidium and/or N,N′-bis(3-(trimethoxysilyl)propyl)formamidinium are formed from the reaction of APTMS and $FA^+$ during vacuum deposition, the reaction proceeds less (or the products fragment more readily) than in the case of AEAPTMS reacting with $FA^+$ during vacuum deposition.

Previously, both our group,[15,19,20,28] and others,[33] have proposed the amino-silane passivators can passivate halide vacancies (undercoordinated $Pb^{2+}$ sites), and that protonated ammoniums could stabilize A-site vacancies in the perovskite surface. Evidence suggests that thin silane treatments also help reduce contact-induced recombination that occurs when the perovskite is in contact with an extraction layer or electrical contact. However, the performance improvements of AEAPTMS, which clearly reacts with $FA^+$ in the perovskite, suggest additional mechanisms. Taddei et al. proposed a lower-dimensional perovskite structure containing the 4,5-dihydroimidazolium cation can form upon reaction with sufficient concentrations of ethylenediamine,[45] while Luther and coworkers have recently proposed that 1D phases are formed by the reaction products of simple alkylamines and diamines with $FA^+$.[47] Presumably these phases form heterojunctions at the surface, which would both push carriers away from surface defects, as well as provide for some electronic decoupling from the electrodes/transport layers.

Finally, we also compare the penetration of the AEAPTMS and APTMS into the perovskite films by tracking the depth (sputter-time) dependence of Si containing fragments. Fig. 4e shows that the initial Si signal is slightly higher for the APTMS-treated films compared to the AEAPTMS-treated films, possibly due to the higher vapor pressure of APTMS leading to faster deposition. Furthermore, the APTMS-treated ToF-SIMS depth profile indicates that Si penetrates deeper into the perovskite bulk compared to the AEAPTMS-treated depth profile with the same treatment duration. We speculate that the higher reactivity of AEAPTMS with $FA^+$, as observed by solution NMR, and bulkier reaction product formed, could help restrict penetration of AEAPTMS into the film, better confining the changes to the surface layer, which could explain the wider processing window that we observe here for AEAPTMS.

In summary, we investigated the effects of APTMS and AEAPTMS surface treatments on halide perovskites. While both amino-silanes can effectively passivate the perovskite surface under optimized conditions, we find that AEAPTMS has a wider processing window, and yields the highest solar cell efficiency, with the most notable improvements observed in the $V_{OC}$ and FF under optimized conditions. We highlight the tunable balance between beneficial passivation and detrimental effects, namely electrical insulation and loss of 3D perovskite phase, influenced by amino-silane treatment duration and conditions.

Using a combination of NMR spectroscopy and depth-resolved ToF-SIMS, we demonstrate that $FA^+$ cations chemically react with these amino-silanes. Overall, our results provide new insight into the interfacial chemistry between amino-silanes and halide perovskites and contribute to the growing evidence that reactivity of $FA^+$ with amine-containing treatments, including amine-functionalized silanes, is a general phenomenon that underpins many additive and passivation approaches. This work also underscores the importance of optimizing treatment conditions to maximize passivation benefits while minimizing adverse effects, an essential consideration for the advancement of halide perovskite semiconductors and the development of robust passivation strategies.

**Supporting Information**

Additional experimental procedures, materials, methods, Tauc plot, XRD, photoluminescence, solar cell data, mass spectra, NMR, schematics of chemical reactions, ToF-SIMS (PDF).

**Author Contributions**

Z.H. and F.A. contributed equally to this work.

**Notes**

The authors declare no competing interests.

**Acknowledgments**

This paper is based primarily on work originally supported by the Office of Naval Research Award (N00014-20-1-2587). The authors acknowledge the use of facilities and instruments at the Photonics Research Center (PRC) at the Department of Chemistry, University of Washington; Research Training Testbed (RTT), part of the Washington Clean Energy Testbeds system. Part of this work is conducted at the Molecular Analysis Facility (MAF), a National Nanotechnology Coordinated Infrastructure site at the University of Washington which is supported in part by funds from the National Science Foundation (awards NNCI-2025489, NNCI-1542101), the Molecular Engineering & Sciences Institute. NMR and mass spectrometry studies were conducted as part of the Center for Hybrid Organic Inorganic Semiconductors for Energy (CHOISE) an Energy Frontier Research Center funded by the Office of Basic Energy Sciences, Office of Science within the U.S. Department of Energy. An Advion Compact Mass Spectrometer used to support the synthetic work at the University of Colorado was purchased using a DURIP equipment grant from the Office of Naval Research (N00014-23-1-2058). We also acknowledge Dr. Fangyuan Jiang at the University of Washington for useful scientific discussions.